\documentclass[11]{article}
\usepackage[]{inputenc}
\usepackage{esvect}
\usepackage[]{cite}
\usepackage{amsmath}
\usepackage{longtable}
\usepackage{fouriernc}
\usepackage[colorlinks=true,citecolor=blue]{hyperref}
\usepackage[left=2.00cm, right=2.00cm, top=2.00cm, bottom=2.00cm]{geometry}

\newcommand{\mt}{\textit{MATHEMATICA}\,}
\newcommand{\ol}{\texttt{Olsson.wl}\,}
\newcommand{\ch}{\texttt{MbConicHull.wl}\,}
\newcommand{\is}{\texttt{Ising.nb}\,}
\newcommand{\bo}{\texttt{Box.nb}\,}
\newcommand{\linefill}{
  {-}\mkern-7mu
  \cleaders\hbox{$\mkern-2mu-\mkern-2mu$}\hfill
  \mkern-7mu{-}%
}
\title{Closed Form Expressions for Certain Improper Integrals of Mathematical Physics }

\author{B. Ananthanarayan \thanks{\href{mailto:anant@iisc.ac.in}{anant@iisc.ac.in}} \hspace{.1cm} Tanay Pathak\thanks{\href{mailto:tanaypathak@iisc.ac.in}{tanaypathak@iisc.ac.in}} \hspace{.1cm} Kartik Sharma\thanks{\href{mailto:kartiksharma@iisc.ac.in}{kartiksharma@iisc.ac.in}}}

\date{Centre for High Energy Physics, Indian Institute of Science,\\ Bangalore-560012, Karnataka, India
}
\begin{document}
\maketitle
\begin{abstract} 
 \noindent We present new closed-form expressions for certain improper integrals of Mathematical Physics such as certain Ising, Box, and Associated integrals. The techniques we employ here include (a) the Method of Brackets and its modifications and suitable extensions to obtain the Mellin-Barnes representation. (b) The evaluation of the resulting Mellin-Barnes representations via the recently discovered Conic Hull method via the automated package \texttt{MBConichulls.wl}. Finally, the analytic continuations of these series solutions are then produced using the automated package \texttt{Olsson.wl}, based on the method of Olsson. Thus, combining all these recent advances allows for closed-form evaluation of the hitherto unknown $B_3(s)$, $B_4(s)$, and related integrals in terms of multivariable hypergeometric functions. Along the way, we also discuss certain complications while using the Original Method of Brackets for these evaluations and how to rectify them. The interesting cases of $C_{5,k}$ are also studied. It is not yet fully resolved for the reasons we discuss in this paper.
\end{abstract}

\setlength{\parindent}{20pt}

\section{Introduction}
Evaluating integrals appearing in various branches of mathematics and mathematical physics continues to be of great importance.  In many instances, one has to resort to numerical methods to evaluate them.  On the other hand, analytical methods are of great importance in their actual evaluation and, in some instances, can also be used to benchmark and test numerical packages.  Of particular note are Feynman Integrals, the basic building blocks of computations in the Standard Model, ineffective theories, and Beyond the Standard Model investigations. As the needs and demands of precision increase, the number of mass scales, as well as the number of loops, also increases very rapidly.  Therefore, recently, a new subject under the Feynman Integral Calculus rubric has come into being, see Smirnov \cite{Smirnov:2006ry}. Furthermore, other techniques such as Mellin-Barnes \cite{Dubovyk:2022obc} and recently developed Conic Hull Mellin Barnes (CHMB) method \cite{Ananthanarayan:2020ncn,Ananthanarayan:2020xpd,Banik:2022quy} are of importance. Another technique that is of importance based on the Ramanujan Master Formula \cite{hardy1999ramanujan} has
become popular is the Method of Brackets \cite{GONZALEZ201050, Gonzalez:2010nm, GonzalezKohlJiuMoll1, quadraticmob,GonzalezJiuMoll2,gonzalez2022mellin,gonzalez2022analytic}. This has been advocated as a method of evaluating Feynman integrals.  However, in practice, there are some technical obstructions to their use \cite{Ananthanarayan:2021not}. All the above said, it is important to understand the basic features of this method.  The present work is an attempt in that direction. It must be pointed out that research of the type reported here is of fundamental and pioneering importance and would be of consequence to developing numerical packages to test their stability and consistency with it. Furthermore, they also help unite disparate subjects of mathematical physics. Such research may be considered as research at the frontier in the year 2023.

A long list of such integrals is already compiled in Gradshyteyn and Ryzik \cite{gradshteyn2014table}. Recently there have been attempts to provide a derivation of a large number of these integrals, specifically the improper integral with limits from $0$ to $\infty$ using the Original Method of Brackets (OMOB)\cite{GONZALEZ201050, Gonzalez:2010nm, GonzalezKohlJiuMoll1, quadraticmob,GonzalezJiuMoll2,gonzalez2022mellin,gonzalez2022analytic}. On the other hand, we would like to emphasize that the results presented in
this work are part of a long series of investigations. Part of our further investigations have been on multi-variate hypergeometric functions, which led us to develop several packages and \texttt{MATHEMATICA} implementation\cite{appellf2,Ananthanarayan:2021yar, Bera:2022eag} many of which are pioneering in their scope and ambit, as well as the potential for application to diverse physical problems.  The present work illustrates this capability, in particular, of combining some rather than using them in isolation.

We turn to the study and evaluation of other improper integrals that appear in Mathematical Physics, such as the Ising and Box integrals. We aim to express them in terms of elegant closed-form expression or in terms of known functions of mathematical physics, especially the hypergeometric functions \cite{srivastava1985multiple,exton1976multiple}.  Several tools have recently been developed to facilitate tasks of symbolic evaluation of these integrals. The recent development of tools and advances in various theoretical treatments has facilitated our results. For instance, the recently proposed solution to the problem of finding the series solution of the $N$-dimensional Mellin-Barnes (MB) representation \cite{Ananthanarayan:2020ncn,Ananthanarayan:2020xpd,Banik:2022quy}, using what has been termed as the Conic Hull Mellin Barnes (CHMB) method. This has also been automated as the \mt package \texttt{MBConichulls.wl} \cite{conichull,Banik:2022bmk}. The series representation hence obtained, in general, can be written as hypergeometric functions or their derivatives. Independently, the issue of finding the analytic continuations (ACs) of the multivariable hypergeometric function using the method of Olsson\cite{Olsson64,appellf2}, which has also been automated as a \mt package \ol \cite{Ananthanarayan:2021yar} have been addressed recently. In this work, we show how these tools, which were primarily directed at solving Feynman integrals, are of sufficient generality to find their use in evaluating the integrals considered here.

We will consider the Ising integrals which have been studied in the Ising model\cite{BAILEY2011741,bailey2006integrals,stan2010recurrences,bailey2007hypergeometric} and also have been in the context of OMOB \cite{Gonzalez:2010nm}. Apart from the evaluation with these newly developed tools, we will also consider certain complications while doing similar evaluations with the OMOB \cite{mobprob}. One of them is the use of regulators for the evaluation of the Ising integrals. This arises in the case of Ising integrals $C_{3,1}$ and $C_{4,1}$. For the case of $C_{4,1}$, it is further complicated due to the use of two regulators, which will give the final result when the proper limiting procedure is applied. However, we point out that such a procedure is complicated and thus use the  Modified Method of Brackets (MMOB)\cite{prausa2017mellin} to get the MB-integral. This MB integral can then be evaluated without introducing such regulators and thus provides an efficient way to deal with these integrals. Using a similar procedure, we attempt to evaluate the elusive $C_{5,k}$ integral. However, we hit a roadblock for the same, as the resulting series does not converge and would require a proper analytic continuation procedure. At present, we find this task beyond the reach of the tools at hand, though we provide a possible way to achieve the same. Yet, such results still shed some light on how these integrals can be evaluated. All the results are provided in the ancillary \mt file \is. 

Box integrals \cite{bailey2007box,bailey2010advances, doi:10.1137/0130003,philip2008distance} are another interesting integrals where such techniques can be applied to get new results. They carry a physical meaning in that they provide the expected distance between two randomly chosen points over the unit $n$-cube. We consider the two special cases of them, namely the $B_{n}(s)$ and the $\Delta_{n}(s)$. We use the same techniques and derive the closed form results for already known $B_{1}(s)$ and $B_{2}(s)$ and new evaluation for $B_{3}(s)$ and $B_{4}(s)$ for general values of $s$. The results are in terms of multivariable hypergeometric function. These evaluations further require an analytic continuation procedure, which has been done using \ol. All the results are provided in the ancillary \mt file \bo. These results for box integrals can then be further used to evaluate the Jellium potential $J_{n}$, which can be related to box integral $B_{n}(s)$\cite{bailey2010advances, BAILEY2007196}. Finally, we give a general MB integral for $B_{n}(s)$, which can be used to find the closed form result for all values of $n$ and $s$ using \ch. With all this, we find new connections between the Box integrals and the multivariable hypergeometric functions. All our calculations rely heavily on \mt as we try to achieve the symbolic results for all the problems.

The paper is structured as follows: In section \eqref{mobrevisit} using an example given in \cite{GonzalezKohlJiuMoll1}, we point out the problem in the OMOB and discuss the alternative to surpass this problem. In section \eqref{ising}, we then proceed to the evaluation of Ising integrals up to $n=4$ while contrasting our method with the previous method to achieve the same in \cite{Gonzalez:2010nm}. In section \eqref{c5ising} we attempt to solve the $C_{5,k}$ integral and point out a general integral $C_{5,k}(\alpha,\beta)$ which gives $C_{5,k}$ as a special case. Though we point out that it is not the final result, a proper analytic continuation procedure is required to get $C_{5,k}$ from it. We then evaluate box integral $B_{n}(s)$ for $n=3,4$ in section \eqref{boxint}. The new results for $\Delta_{n}(s)$ and $J_{n}$ with the above new results are also provided. Finally, we conclude the paper with some conclusions and possible future directions in section \eqref{conclusions}. In appendix \ref{mfiles}, we provide the table for all the \mt files that we give and the packages required, which are available in \href{https://github.com/TanayPathak-17/Improper-Integrals}{GitHub}.

\section{Method of Brackets revisited}\label{mobrevisit}
We will first illustrate the OMOB using a simple example of integral evaluation as given in \cite{GonzalezKohlJiuMoll1}. We will first evaluate the integral by using the OMOB and then describe the difficulty in applying the method directly. We will then propose a possible resolution to carry out evaluations for such cases and then illustrate the alternative method to do the same.\\  We consider the following integral 
\begin{equation}\label{emobeg}
H_1(a, b)=\int_0^{\infty} K_0(a x) K_0(b x)\; \mathrm{d}x
\end{equation}
The integral is introduced to facilitate the evaluation of another integral, which is given by putting $a=b$
\begin{equation}\label{emobeg2}
    H(a)=\int_0^{\infty} K_0^2(a x)\; \mathrm{d} x
\end{equation}
We can express $K_0 (x)$ using the following series expansion:
\begin{equation}
    K_0(ax) = \sum_{n_1} \phi_{n_1} \frac{a^{2n_1}\Gamma(-n_1)}{2^{2n_1 + 1}} x^{2n_1}
\end{equation}
where $\phi_{n} = \frac{(-1)^{n}}{\Gamma(n+1)}$.

This expansion uses a divergent series, and we can express the result in the form of an integral representation as
\begin{equation}
    K_0(bx) = \frac{1}{2} \int_0^{\infty} {\exp} \left(-t - \frac{b^2 x^2}{4t}\right) \frac{\mathrm{d}t}{t}
\end{equation}
Using the OMOB, we get:
\begin{equation}
    K_0(bx) = \sum_{n_2,n_3} \phi_{n_2,n_3} \frac{b^{2n_3}x^{2n_3}}{2^{2n_3+1}} \langle n_2 - n_3 \rangle
\end{equation}
Substituting the bracket series in  Eq.\eqref{emobeg}, we get
\begin{equation}
    H_1(a,b) = \sum_{n_1,n_2,n_3} \phi_{n_1,n_2,n_3} \frac{a^{2n_1}b^{2n_3}\Gamma(-n_1)}{2^{2n_1+2n_3+2}} \langle n_2 - n_3 \rangle \langle 2n_1 + 2n_3 + 1 \rangle
\end{equation}
Now, we need to solve the bracket equations, which involve 2 equations but three variables. Evaluating this we get the following three series, $T_{i}$ where $n_i$ is the free variable:
\begin{align}\label{probsol}
    T_1 = \frac{1}{4a}\sum_n \phi_n \Gamma(-n) \Gamma^2 \left(n + \frac{1}{2}\right) \left(\frac{b}{a}\right)^{2n} \nonumber \\
    T_2 = \frac{1}{4a}\sum_n \phi_n \Gamma(-n) \Gamma^2 \left(n + \frac{1}{2}\right) \left(\frac{b}{a}\right)^{2n} \nonumber \\
    T_3 = \frac{1}{4a}\sum_n \phi_n \Gamma(-n) \Gamma^2 \left(n + \frac{1}{2}\right) \left(\frac{b}{a}\right)^{2n}
\end{align}
Using the rules of the OMOB, all three series of Eq.\eqref{probsol} have to be discarded as they are divergent.

A solution to such a problem, as implemented in \cite{GonzalezKohlJiuMoll1}, is to regularize the singularity. This amounts to modifying the bracket $\langle n_2 - n_3 \rangle \rightarrow \langle n_2 - n_3 +\epsilon \rangle$.
With this modification, when $n_1$ is a free variable, one gets the series that contains $\Gamma(-n)$, which is diverging and is thus discarded. For the other cases, one gets two series with $\epsilon$ parameter (in the form of $\Gamma(-n + \epsilon)$ and $\Gamma(-n - \epsilon)$). In these series, when the proper limiting procedure is done, along with the condition $a=b$ to ease the calculation, they give the result for the integral of Eq.\eqref{emobeg2}. Thus, the original integral of Eq.\eqref{emobeg} we started with still remains elusive, as the calculation is much more involved (the limiting procedure) within this present framework.

An alternative to the above evaluation, free from choosing the regulator and doing the tedious limiting procedure, is to use the MB representation derived using the MMOB\cite{prausa2017mellin}. Using it, we get the following MB representation for the integral given by Eq.\eqref{emobeg}
\begin{equation}
  H_1(a, b)= \frac{1}{4} \int\limits_{c - i \infty} ^{c + i \infty} \frac{\mathrm{d}z}{2 \pi i}\;a^{-2 z-1} b^{2 z} \Gamma (-z)^2 \Gamma \left(\frac{1}{2} (2 z+1)\right)^2
\end{equation}
The above MB integral can be readily evaluated in \mt to give the following result
\begin{equation}
   H_1(a, b)= \frac{\pi  \sqrt{\frac{a^2}{b^2}} K\left(1-\frac{a^2}{b^2}\right)}{2 a}
\end{equation}
where $K(x)$ is the complete elliptic integral of the first kind. Thus, we get the value of the original integrals, Eq.\eqref{emobeg} we started with.\\
For the special case of $a=b$, using $K(0)=\frac{\pi}{2}$ we get 
\begin{equation}
    H_1(a,a)= H(a)= \frac{\pi^2}{4 a}
\end{equation}
So, we see that for the simple cases, too, using the MB representation to evaluate these integrals provides an efficient way to evaluate these integrals.

\section{Ising integrals}\label{ising}
In this section, we will analyze the integrals of the ``Ising class''. Ising models are extensively used to study the statistical nature of ferromagnets \cite{orrick2001susceptibility,PhysRevB.13.316,zenine2005square}. The model accounts for the magnetic dipole moments of the spins. The $n$ - dimensional integrals are denoted by $C_n, D_n, E_n$, where $D_n$ is found in the magnetic susceptibility integrals essential to the Ising calculations.
\begin{equation}
    D_n = \frac{4}{n!} \int_0 ^{\infty} \cdots \int_0^{\infty} \frac{  \prod_{i<j} \left(\frac{u_i - u_j}{u_i + u_j}\right)^2}{(\sum_{j=1} ^n (u_j + 1/u_j))^2} \frac{\mathrm{d}u_1}{u_1} \cdots \frac{\mathrm{d}u_n}{u_n}
\end{equation}
The integral $D_n$ provides great insights into the symmetry breaking at low-temperature phases and finds great use in quantum field theories and condensed matter physics. However, it is difficult to evaluate these integrals computationally and analytically. On the other hand, the $C_n$ ($C_n = C_{n,1}$) class integrals, which are closely related to the $D_n$ class, are easier to tackle and can produce closed-form expressions.

The general Ising integrals $C_{n,k}$ is defined as
\begin{equation}
     C_{n,k} = \frac{4}{n!}  \int_0 ^{\infty} \cdots \int_0 ^{\infty} \frac{1}{(\sum_{j=1} ^n (u_j + 1/u_j))^{k+1}} \frac{\mathrm{d}u_1}{u_1} \cdots \frac{\mathrm{d}u_n}{u_n}
\end{equation}
    The above expression can also be expressed as the moments of power of Bessel Function $K_0$ as
\begin{equation}
    C_{n,k} = \frac{2^{n-k+1}}{n! \; k!} c_{n,k} := \frac{2^{n-k+1}}{n! \; k!} \int_0 ^{\infty} t^k K_0 ^n (t) \;\mathrm{d}t
\end{equation}
We will now analyze the special case of the $C_{n,k}$ family with $k=1$ using the Method of Brackets \cite{GONZALEZ201050, Gonzalez:2010nm, bailey2006integrals} and Mellin-Barnes representations. After this, each general integral with $C_{n,k}$ will be treated using the same procedure. The $C_{1,k}$ and $C_{2,k}$ integrals are easily tractable, and the results for them have been given just for completeness' sake. The problem occurs when one considers $C_{n,k}$ for $n \geq 3$.
Below, we use the MMOB \cite{prausa2017mellin} and show that for the evaluation of the integrals requiring the use of regulators, it is better to use the MMOB and solve the corresponding integral using the CHMB method. The main utility of the method is that the limiting procedure is automatically taken care of while finding the residue in the case of CHMB, which is at times difficult, especially when there is more than 1 regulator, as in the case of $C_{4,k}$.
\subsection{$C_{1,k}$}
For $n = 1$, we have
\begin{equation}
    C_{1,k} = \frac{4}{1!} \int_0 ^{\infty} \frac{1}{(u_1 + 1/u_1)^{k+1}} \frac{\mathrm{d}u_1}{u_1}
\end{equation}
The integral can simply be evaluated to give the general closed form:
\begin{equation}
    C_{1,k} = \frac{\sqrt{\pi } 2^{1-k} \Gamma \left(\frac{k+1}{2}\right)}{\Gamma \left(\frac{k}{2}+1\right)}
\end{equation}

\subsection{$C_{2,k}$}
For $n=2$, we get:
\begin{equation}
     C_{2,k} = \frac{4}{2!}  \int_0 ^{\infty} \int_0 ^{\infty} \frac{1}{(u_1 + 1/u_1 + u_2 + 1/u_2)^{k+1}} \frac{\mathrm{d}u_1}{u_1} \frac{\mathrm{d}u_2}{u_2}
\end{equation}
This evaluation using the MOB, for $k=1$, gives:
\begin{equation}
    C_{2,1} = 1
\end{equation}
The integral for the general value of $k$ can also be evaluated to give the following closed form:
\begin{equation}
    C_{2,k} = \frac{\Gamma \left(\frac{k}{2}+\frac{1}{2}\right)^4}{\Gamma (k+1)^2}
\end{equation}
\subsection{$C_{3,k}$ and $C_{3,k}(\alpha,\beta,\gamma)$}
For $k=1$, we get:
\begin{equation}
     C_{3,1} = \frac{4}{3!}  \int_0 ^{\infty}  \int_0 ^{\infty} \int_0 ^{\infty} \frac{1}{(u_1 + 1/u_1 + u_2 + 1/u_2 + u_3 + 1/u_3)^2} \frac{\mathrm{d}u_1}{u_1} \frac{\mathrm{d}u_2}{u_2} \frac{\mathrm{d}u_3}{u_3}
\end{equation}
We will illustrate the problem encountered in OMOB by writing the bracket series for the generalized case $C_{3,k}$.\\
The following form of the integrand is motivated to maximize the number of brackets series in the expansion, which in turn reduces the number of variables:
\begin{equation}
    C_{3,k} = \frac{2}{3}  \int_0 ^{\infty}  \int_0 ^{\infty} \int_0 ^{\infty} \frac{(u_1 u_2 u_3)^k}{(u_1 u_2 u_3 (u_1 + u_2) + u_3 (u_1 + u_2)+u_1 u_2 u_3^2 + u_1 u_2)^{k+1}} \mathrm{d}u_1 \mathrm{d}u_2 \mathrm{d}u_3
\end{equation}
Expanding the denominator using the rules of MOB, 
\begin{equation}
    \sum_{\{n\}} \phi_{\{n\}} (u_1 u_2)^{n_1 + n_3+ n_4} u_3^{n_1+n_2+2n_3} (u_1 + u_2)^{n_1+n_2} \frac{\langle k+1 + n_1+ n_2+ n_3 + n_4 \rangle}{\Gamma(k+1)}
\end{equation}
Now, $(u_1 + u_2)^{n_1+n_2}$ has to be further expanded as:
\begin{equation}
    (u_1 + u_2)^{n_1+n_2} = \sum_{n_5,n_6} \phi_{n_5,n_6} u_1^{n_5} u_2^{n_6} \frac{\langle -n_1 -n_2 + n_5 + n_6 \rangle}{\Gamma(-n_1-n_2)}
\end{equation}
Combining the expansions, the $C_{3,k}$ integral takes the form:
\begin{align}
    C_{3,k} &= \frac{2}{3 \Gamma(k+1)} \sum_{\{n\}} \phi_{\{n\}} \frac{\langle -n_1 -n_2 + n_5 + n_6 \rangle}{\Gamma(-n_1-n_2)}  \\ \nonumber
    & \;\;\;\times \langle k+1 + n_1+ n_3 + n_4 + n_5\rangle \langle k+1 + n_1+ n_3 + n_4 + n_6\rangle \\ \nonumber
    & \;\;\;\times \langle k+1+n_1+n_2+2n_3\rangle \langle k+1 + n_1+ n_2+ n_3 + n_4 \rangle
\end{align}
Now, the rules of MOB demand that we solve the linear equations of the brackets, but that poses the problem of giving rise to divergent terms like $\Gamma(-n)$ and renders the whole procedure useless. To solve the issue, it is suggested to introduce regulators. For the case of $C_{3,k}$, one regulator is enough. In particular, $\epsilon \,(\rightarrow 0)$ is introduced in the bracket as $\langle k+1+n_1+n_2+2n_3\rangle \rightarrow \langle k+1+n_1+n_2+2n_3 +\epsilon \rangle$ which mimics the effect of introducing a factor of $u_3^{\epsilon}$ in the integrand. Now, with this ``new'' bracket series, the divergent terms take the form of $\Gamma(-n-\epsilon)$ and are easier to work with. In the regime of OMOB, one requires the expansion of $\Gamma(x)$ around integers to deal with the problem, which increases the complexity of the task.

As $n$ increases, the number of regulators increases monotonically and complicates the limiting procedure. On the other hand, MMOB doesn't call for any regulators and is very computationally friendly.
Using the MMOB in the above bracket series, we get the following MB representation for the $C_{3,1}$:
\begin{equation}
    C_{3,1} = \frac{1}{3} \int\limits_{c - i \infty} ^{c + i \infty} \frac{\mathrm{d}z}{2 \pi i}\; \frac{\Gamma(-z)^4 \; \Gamma(1+z)^2}{\Gamma(-2z)}
\end{equation}
This evaluates to
\begin{equation}
    C_{3,1} = \frac{2}{27} \left(6 i \sqrt{3} \left(\text{Li}_2\left(\frac{1}{4}-\frac{i \sqrt{3}}{4}\right)-\text{Li}_2\left(\frac{i \sqrt{3}}{4}+\frac{1}{4}\right)\right)+\pi  \sqrt{3} \log (4)-\psi ^{(1)}\left(\frac{1}{3}\right)+\psi ^{(1)}\left(\frac{2}{3}\right)\right)
\end{equation}
where $\psi^{(1)}$ is the polygamma function of order 1. \\

The generalized integral $C_{3,k}$ can be similarly obtained using the MMOB to give the following MB representation:
\begin{equation}
  C_{3,k}= \frac{1}{3 \Gamma(k+1)}\int\limits_{c - i \infty} ^{c + i \infty} \frac{\mathrm{d}z}{2\pi i} \frac{\Gamma (-z)^4 \; \Gamma \left(\frac{1}{2} (k+2 z+1)\right)^2}{\;\Gamma (-2 z)}
\end{equation}
The above integral can be evaluated to give
\begin{equation}
    C_{3,k} = \frac{2}{3 \, k!} \sqrt{\pi }\, G_{3,3}^{2,3}\left(\frac{1}{4} \left|
\begin{array}{c}
 1,1,1 \\
 \frac{k+1}{2},\frac{k+1}{2},\frac{1}{2} \\
\end{array}
\right. \right)
\end{equation}
where $G_{p, q}^{m, n}$ is the Meijer-G function, which is given by following Mellin-Barnes representation \cite{bateman1953higher}
\begin{equation*}
G_{p, q}^{m, n}\left(\begin{array}{c}
a_1, \ldots, a_p \\
b_1, \ldots, b_q
\end{array} \mid z \right)=\frac{1}{2 \pi i} \int_L \frac{\prod_{j=1}^m \Gamma\left(b_j-s\right) \prod_{j=1}^n \Gamma\left(1-a_j+s\right)}{\prod_{j=m+1}^q \Gamma\left(1-b_j+s\right) \prod_{j=n+1}^p \Gamma\left(a_j-s\right)} z^s \mathrm{d} s
\end{equation*}
A further generalization of $C_{3,k}$ integral namely $C_{3,k}(\alpha,\beta,\gamma)$ is given in \cite{Gonzalez:2010nm} where the following integral is considered
\begin{equation}
    C_{3, k}(\alpha, \beta, \gamma)=\int_0^{\infty} \int_0^{\infty} \int_0^{\infty} \frac{x^{\alpha-1} y^{\beta-1} z^{\gamma-1}}{(x+1/x + y+1/y + z+1/z)^{k+1}} \mathrm{d}x \mathrm{d}y \mathrm{d}z
\end{equation}
Using the MMOB, we get the following MB representation
\begin{equation}
   C_{3, k}(\alpha, \beta, \gamma)= \frac{1}{3 \Gamma(k + 1)}\int\limits_{c - i \infty} ^{c + i \infty} \frac{\mathrm{d}z}{2 \pi i}\;\frac{\Gamma (-z) \Gamma (-z+\alpha -1) \Gamma (-z-\beta +1) \Gamma (-z+\alpha -\beta ) \Gamma \left(\frac{1}{2} (k+2 z-\alpha +\beta -\gamma +2)\right) \Gamma \left(\frac{1}{2} (k+2 z-\alpha +\beta +\gamma )\right)}{\Gamma (-2 z+\alpha -\beta )}
\end{equation}
The result is given in the \mt file and is found to be :
\begin{align}
    \nonumber
    &=-\frac{1}{3 k!}\pi ^{3/2} \csc (\pi  \gamma ) 2^{-\gamma -k-1} \Big(4^{\gamma } \Gamma \left(\frac{1}{2} (k-\alpha -\beta -\gamma +4)\right) \Gamma \left(\frac{1}{2} (k+\alpha -\beta -\gamma +2)\right) \Gamma \left(\frac{1}{2} (k-\alpha +\beta -\gamma +2)\right) \Gamma \left(\frac{1}{2} (k+\alpha +\beta -\gamma )\right) \\ \nonumber
    & \times \, _4\tilde{F}_3\left(\frac{1}{2} (k+\alpha +\beta -\gamma ),\frac{1}{2} (k-\alpha -\beta -\gamma +4),\frac{1}{2} (k+\alpha -\beta -\gamma +2),\frac{1}{2} (k-\alpha +\beta -\gamma +2);\frac{1}{2} (k-\gamma +2),\frac{1}{2} (k-\gamma +3),2-\gamma ;\frac{1}{4}\right) \\ \nonumber
    & -4 \Gamma \left(\frac{1}{2} (k-\alpha -\beta +\gamma +2)\right) \Gamma \left(\frac{1}{2} (k+\alpha -\beta +\gamma )\right) \Gamma \left(\frac{1}{2} (k-\alpha +\beta +\gamma )\right) \Gamma \left(\frac{1}{2} (k+\alpha +\beta +\gamma -2)\right) \\ 
    & \times \, _4\tilde{F}_3\left(\frac{1}{2} (k-\alpha +\beta +\gamma ),\frac{1}{2} (k+\alpha +\beta +\gamma -2),\frac{1}{2} (k-\alpha -\beta +\gamma +2),\frac{1}{2} (k+\alpha -\beta +\gamma );\frac{k+\gamma }{2},\frac{1}{2} (k+\gamma +1),\gamma ;\frac{1}{4}\right) \Big) 
\end{align}

where 
\begin{equation*}
_{4}\tilde{F}_{3}(a_{1},a_{2},a_{3},a_{4};b_{1},b_{2},b_{3};x) = \frac{1}{\Gamma(b_{1})\Gamma(b_{2})\Gamma(b_{3})} \sum_{n=0}^{\infty} \frac{(a_1)_n(a_2)_n (a_3)_n (a_4)_n}{(b_1)_n (b_2)_n (b_3)_n} \frac{x^n}{n !}
\end{equation*}
 and $(a)_n = \frac{\Gamma(a+n)}{\Gamma(a)}$ is the Pochhammer symbol. 
\subsection{$C_{4,k}$ and $C_{4,k}(\alpha,\beta,\gamma,\delta)$}
For $k=1$:
\begin{equation}
     C_{4,1} = \frac{4}{4!}  \int_0 ^{\infty} \int_0 ^{\infty}  \int_0 ^{\infty} \int_0 ^{\infty} \frac{1}{(u_1 + 1/u_1 + u_2 + 1/u_2 + u_3 + 1/u_3 + u_4 + 1/u_4)^2} \frac{\mathrm{d}u_1}{u_1} \frac{\mathrm{d}u_2}{u_2} \frac{\mathrm{d}u_3}{u_3} \frac{\mathrm{d}u_4}{u_4}
\end{equation}
If one proceeds with the OMOB as in the case of $C_{3,1}$, one is now required to use 2 regulators, namely $\epsilon$ and $A$ \cite{Gonzalez:2010nm}. The result for $C_{4,1}$ is then obtained by taking the limit $\epsilon \rightarrow 0$, A$\rightarrow 1$. The use of two regulators significantly complicates the task of doing the limiting procedure. So, we again proceed with the use of the MMOB.
Using the MOB, we get the following MB representation for $C_{4,1}$:
\begin{equation}
    C_{4,1} = \frac{1}{12} \int\limits_{c - i \infty} ^{c + i \infty} \frac{\mathrm{d}z}{2 \pi i}\; \frac{\Gamma(-z)^4 \; \Gamma(1+z)^4}{\Gamma(-2z) \; \Gamma(2+2z)}
\end{equation}
This can be evaluated to give
\begin{equation}
    C_{4,1} = \frac{7 \zeta (3)}{12}
\end{equation}
The general case for $n=4$ can be simplified to the following MB representation:
\begin{equation}
  C_{4,k}=  \frac{1}{12 \Gamma (k+1) \;}\int\limits_{c - i \infty} ^{c + i \infty} \frac{\mathrm{d}z}{2 \pi i}\; \frac{\Gamma (-z)^4 \;\Gamma \left(\frac{k+1}{2}+z\right)^4}{\Gamma (-2 z) \; \Gamma (k+2 z+1)}
\end{equation}
This can be evaluated to give the closed-form expression:
\begin{equation}
    C_{4,k} = \frac{\pi \;  2^{-k-1}}{3 \Gamma (k+1)}  G_{4,4}^{3,3}\left(1\left|
\begin{array}{c}
 1,1,1,\frac{k+2}{2} \\
 \frac{k+1}{2},\frac{k+1}{2},\frac{k+1}{2},\frac{1}{2} \\
\end{array}
\right.\right)
\end{equation}
The given expression is of particular interest, as seen from its values when evaluated for any odd values of $k$. When $C_{4,k}$ is evaluated for any odd $k$, it takes the form of $a \zeta(3) + b$ function, where $a$ and $b$ are some rational numbers. Some of the values are provided for reference in Table~\ref{c4table}.

\begin{table}[h]
\centering
\begin{tabular}{|c|c|}
\hline
 $k$ & $C_{4,k}$  \\ \hline
 &  \\
 0 & $\frac{1}{6} \pi  G_{4,4}^{3,3}\left(1\left|
\begin{array}{c}
 1,1,1,1 \\
 \frac{1}{2},\frac{1}{2},\frac{1}{2},\frac{1}{2} \\
\end{array}
\right.\right)$ \\
 &  \\ \hline
 &  \\
 1 & $\frac{7 \zeta (3)}{12}$ \\
 &  \\ \hline
 &  \\
 2 & $\frac{1}{48} \pi  G_{4,4}^{3,3}\left(1\left|
\begin{array}{c}
 1,1,1,2 \\
 \frac{3}{2},\frac{3}{2},\frac{3}{2},\frac{1}{2} \\
\end{array}
\right.\right)$ \\
 &  \\ \hline
 &  \\
 3 & $\frac{7 \zeta (3)-6}{1152}$ \\
 &  \\ \hline
 &  \\
 4 & $\frac{1}{2304} \pi  G_{4,4}^{3,3}\left(1\left|
\begin{array}{c}
 1,1,1,3 \\
 \frac{5}{2},\frac{5}{2},\frac{5}{2},\frac{1}{2} \\
\end{array}
\right.\right)$ \\
 &  \\ \hline
 &  \\
 5 & $\frac{49 \zeta (3)-54}{368640}$ \\
 &  \\ \hline
 &  \\
 6 & $\frac{1}{276480}\pi  G_{4,4}^{3,3}\left(1\left|
\begin{array}{c}
 1,1,1,4 \\
 \frac{7}{2},\frac{7}{2},\frac{7}{2},\frac{1}{2} \\
\end{array}
\right.\right)$ \\
 &  \\ \hline
 &  \\
 7 & $\frac{63 \zeta (3)-74}{15482880}$  \\
 &  \\ \hline
\end{tabular}
\caption{Values of $C_{4,k}$ for $k=0, \cdots, 7$}
\label{c4table}
\end{table}
A further generalization of $C_{4,k}$ integral namely $C_{4,k}(\alpha,\beta,\gamma,\delta)$ can be considered as follows:
\begin{equation}
    C_{4, k}(\alpha, \beta, \gamma,\delta)=\int_0^{\infty}\int_0^{\infty} \int_0^{\infty} \int_0^{\infty} \frac{x^{\alpha-1} y^{\beta-1} z^{\gamma-1} w^{\delta-1} }{(x+1/x + y+1/y + z+1/z +w+1/w)^{k+1}}\mathrm{d} x \mathrm{d} y \mathrm{d} z \mathrm{d}w
\end{equation}
Using the MMOB, we get the following MB representation
\begin{align}
   C_{4, k}(\alpha, \beta, \gamma)= \frac{1}{12  \Gamma(k+1)}&\int\limits_{c - i \infty} ^{c + i \infty} \frac{\mathrm{d}z}{2 \pi i}\; \frac{\Gamma (-z) \Gamma (-z+\gamma -1) \Gamma (-z-\delta +1) \Gamma (-z+\gamma -\delta ) \Gamma \left(\frac{1}{2} (k+2 z-\alpha -\beta -\gamma +\delta +3)\right)  }{12 \Gamma (k+1) \Gamma (-2 z+\gamma -\delta ) \Gamma \left(\frac{1}{2} (k+2 z-\alpha -\beta -\gamma +\delta +3)+\frac{1}{2} (k+2 z+\alpha +\beta -\gamma +\delta -1)\right)} \nonumber \\
 & \times \Gamma \left(\frac{1}{2} (k+2 z+\alpha -\beta -\gamma +\delta +1)\right) \Gamma \left(\frac{1}{2} (k+2 z-\alpha +\beta -\gamma +\delta +1)\right)  \Gamma \left(\frac{1}{2} (k+2 z+\alpha +\beta -\gamma +\delta -1)\right)
\end{align}
The above integral can also be evaluated as before, and the solution has been provided in the accompanying \mt file \texttt{Ising.nb}.

We end this section by noting that given an integral, the evaluation of its MB representation obtained using the MMOB\cite{prausa2017mellin} is more efficient than using the OMOB and its rules to evaluate the same. The regulators and the limiting procedure in the OMOB are automatically taken care of in the evaluation of MB integrals while evaluating the residue. Alternatively, this suggests that one can try to find a better rule that concerns the elimination of the bracket for the OMOB so that one does not require regulators and the result is obtained with their use.

\section{An attempt at $C_{5,k}$}\label{c5ising}
Using the machinery developed so far, we now attempt to evaluate the $C_5$ integral in the same spirit. Using the MOB, we get the following MB representation for $C_{5,k}$
\begin{align}\label{c5ori}
   C_{5,k}= \frac{1}{60 \Gamma (k+1) } \int\limits_{c_1 - i \infty} ^{c_1 + i \infty} \frac{\mathrm{d}z_1}{2 \pi i}\; \int\limits_{c_2 - i \infty} ^{c_2 + i \infty} \frac{\mathrm{d}z_2}{2 \pi i}\; \frac{\Gamma \left(-z_1\right){}^4 \Gamma \left(-z_2\right){}^4 \Gamma \left(\frac{1}{2} \left(k+2 z_1+2 z_2+1\right)\right){}^2}{\Gamma \left(-2 z_1\right) \Gamma \left(-2 z_2\right)}
\end{align}
Evaluation of the above integral, when done directly using the \texttt{MBConicHulls.wl}, would result in the divergent series. A suitable way to approach such evaluation would be by taking two parameters that serve as the variables for the series that appear and then evaluating the results with these parameters. For the $C_{5,k}$ integral we have the following evaluation
\begin{align}\label{c5ab}
   C_{5,k}(\alpha,\beta)= \frac{1}{60 \Gamma (k+1) } \int\limits_{c_1 - i \infty} ^{c_1 + i \infty} \frac{\mathrm{d}z_1}{2 \pi i}\; \int\limits_{c_2 - i \infty} ^{c_2 + i \infty} \frac{\mathrm{d}z_2}{2 \pi i}\; (\alpha)^{z_1}(\beta)^{z_2}\frac{\Gamma \left(-z_1\right){}^4 \Gamma \left(-z_2\right){}^4 \Gamma \left(\frac{1}{2} \left(k+2 z_1+2 z_2+1\right)\right){}^2}{\Gamma \left(-2 z_1\right) \Gamma \left(-2 z_2\right)}
\end{align}
We notice that the integral Eq.\eqref{c5ab} has a more general structure than the integral Eq.\eqref{c5ori} with the introduction of the two parameters. The $C_{5,k}$ can be obtained by putting $\alpha=\beta=1$.
The evaluation of the Eq.\eqref{c5ab} has been done in the accompanying \mt file \texttt{Ising.nb}.

We also note that though we have a result for integral \eqref{c5ab}, the result is not convergent for the value of interest $\alpha=\beta=1$. Proper analytic continuation techniques have to be used to achieve this goal. At present, with the form of series that we obtain, the task is not achievable using \ol. With the form of series at hand, we believe that it can be written as a derivative of some hypergeometric function. Then \ol can be used to find the analytic continuations of this hypergeometric function so that it converges for $\alpha=\beta=1$, and then the derivative can be performed to get the final result. 

\section{Box Integrals}\label{boxint}

For dimension $n$, we define the box integral as the expected distance from a fixed point $\boldsymbol{q}$ (can be origin also) of point $\boldsymbol{r}$ chosen randomly and independently over the unit $n$-cube, with parameter $s$,
\begin{equation}
    B_n(s) = \int_0 ^1 \cdots\int_0 ^1 \Big((r_1)^2 + \cdots + (r_n)^2 \Big)^{s/2} \, \mathrm{d}r_1 \cdots \mathrm{d}r_n
\end{equation}
\begin{equation}\label{deltaint}
    \Delta_n(s) = \int_0 ^1 \cdots\int_0 ^1 \Big((r_1 - q_1)^2 + \cdots + (r_n - q_n)^2 \Big)^{s/2} \, \mathrm{d}r_1 \cdots \mathrm{d}r_n \mathrm{d}q_1 \cdots \mathrm{d}q_n 
\end{equation}
For certain special values of parameter $s$, the above integrals give the following interpretation:
\begin{enumerate}
    \item $B_n (1)$: It gives the expected distance from the origin for a random point of the $n$-cube.
    \item $\Delta_n (1)$: It gives the expected distance between two random points of the $n$-cube.
\end{enumerate}
Due to the physical significance of the box integrals and their use in the electrostatic potential calculations, we wanted to evaluate these integrals and give closed-form expressions using the Method of Brackets that has been implemented throughout the paper.\\

Using the quadrature formulae for all complex powers \cite{bailey2007box, bailey2010experimental, BAILEY2007196, bailey2011high, bailey2010advances}, we use the functions:
\begin{equation}
    b(u) = \int_0^{1} e^{-u^2 x^2} \mathrm{d}x = \frac{\sqrt{\pi} \, \text{erf}(u)}{2 u}
\end{equation}
\begin{equation}
    d(u) = \int _0^1\int _0^1e^{-u^2 (x-y)^2} \mathrm{d}y \,\mathrm{d}x = \frac{\sqrt{\pi } \, u \, \text{erf}(u)+e^{-u^2}-1}{u^2}
\end{equation}
which gives us the relation:
\begin{equation}
    B_n(s) = \frac{2 }{\Gamma \left(-s/2\right)} \int_0^{\infty } u^{-s-1} b^n(u) \, \mathrm{d}u
\end{equation}
\begin{equation}
    \Delta_n(s) = \frac{2 }{\Gamma \left(-s/2\right)} \int_0^{\infty } u^{-s-1} d^n(u) \, \mathrm{d}u
\end{equation}
\subsection{$B_n(s)$}
Now, for the method of brackets to be operational, we need integrals of the form with limits from $0$ to $\infty$.
We need to make an Euler substitution. The following substitution has been found to be the most efficient:
\begin{equation}
    x \rightarrow \frac{a}{1+a}
\end{equation}
which makes the integral
\begin{equation}
    b(u) = \int_0^{1} e^{-u^2 x^2} \mathrm{d}x = \int_0^{\infty} e^{-u^2 \left( \frac{a}{1+a} \right)^2} \frac{1}{(1+a)^2} \,\mathrm{d}a
\end{equation}
\begin{equation}
    b(u) = \int_0^{\infty} \sum_{n=0} ^{\infty} \, \frac{1}{n!}\,\Bigg(\frac{-u^2 a^2}{(1+a)^2}\Bigg)^n \frac{1}{(1+a)^2} \,\mathrm{d}a
\end{equation}

Substituting this back in $B_n(u)$ and applying MMOB, it is obtained that $B_n(s)$ has a pole at $s = -n$ and we finally get:
\begin{equation}
    B_1(s) = \frac{1}{s+1} , s \neq - 1
\end{equation}
\begin{equation}\label{b2sol}
    B_2(s) = \frac{2}{s+2} \, _2F_1\left(\frac{1}{2},-\frac{s}{2};\frac{3}{2};-1\right) , s \neq - 2
\end{equation}
The first two cases were easy to handle. The first non-trivial evaluation is that of $B_3(s)$. We found two different results for the same by using two different methods. Firstly we consider the following representation of $B_3$ \cite{bailey2010advances}:
\begin{equation}
B_3(s)=\frac{3}{3+s} C_{2,0}(s, 1)=\frac{6}{(3+s)(2+s)} \int_0^{\pi / 4}\left(\left(1+\sec ^2 t\right)^{s / 2+1}-1\right)
\end{equation}
The above can interestingly be evaluated in \textit{MATHEMATICA} using \texttt{Integrate} command. Using it, we get the following evaluation for the $B_3(s)$ integral
\begin{align}\label{b3result1}
     B_3(s) &= \frac{6}{(s+2) (s+3)} \, \Bigg ( i F_1\left(1;\frac{1}{2},-\frac{s}{2};2;2,-2\right)-\frac{2^{\frac{s+1}{2}}}{s+1} F_1\left(\frac{1}{2} (-s-1);-\frac{1}{2},-\frac{s}{2};\frac{1-s}{2};\frac{1}{2},-\frac{1}{2}\right) \nonumber \\
        & -i \, _2F_1\left(1,-\frac{s}{2};\frac{3}{2};-1\right)+2^{s/2} \, _2F_1\left(\frac{1}{2},-\frac{s}{2};\frac{3}{2};-\frac{1}{2}\right) -\frac{\sqrt{\pi }}{4 \Gamma \left(1-\frac{s}{2}\right)} \, _2F_1\left(-\frac{s}{2}-\frac{1}{2},-\frac{s}{2};1-\frac{s}{2};-1\right) \Gamma \left(-\frac{s}{2}-\frac{1}{2}\right) -\frac{\pi }{4} \Bigg)
\end{align}
where $F_1(a;b_1,b_2;c;x,y)$ is the Appell $F_1$ function which is defined for $|x| < 1 \wedge |y| < 1$ as:
\begin{equation}
    F_1(a;b_1,b_2;c;x,y) = \sum_{m,n=0} ^{\infty} \frac{(a)_{m+n} (b_1)_m (b_2)_n}{(c)_{m+n} m! n!} x^m y^n
\end{equation}
where $(q)_n$ is the Pochhammer symbol.

Eq.\eqref{b3result1} requires the evaluation of the Appell $F_1$ outside its region of convergence. Such evaluation requires the use of analytic continuation of $F_1$, which has been done by Olsson \cite{olsson1964integration}. \\
Though we got the result using \mt, it doesn't provide many insights so as to aid the computations of other $B_{n}(s)$. So we proceed to a more systematic evaluation of the $B_{3}(s)$ so that the results can be generalized to other values of $n$.
Using the MMOB\cite{prausa2017mellin} we get the following Mellin-Barnes integral for the $B_{3}(s)$
\begin{equation}
    B_{3}(s)=  \int\limits_{c_1 - i \infty} ^{c_1 + i \infty}\, \int\limits_{c_2 - i \infty} ^{c_2 + i \infty} \frac{\Gamma \left(-z_1\right) \Gamma \left(-z_2\right) \Gamma \left(2 z_1+1\right) \Gamma \left(2 z_2+1\right) \Gamma \left(s-2 z_1-2 z_2+1\right) \Gamma \left(-\frac{s}{2}+z_1+z_2\right)}{\Gamma \left(-\frac{s}{2}\right) \Gamma \left(2 z_1+2\right) \Gamma \left(2 z_2+2\right) \Gamma \left(s-2 z_1-2 z_2+2\right)} \frac{\mathrm{d}z_{2}}{2 \pi i} \frac{\mathrm{d}z_{1}}{2 \pi i}
\end{equation}
We evaluate the above integral using the \texttt{MBConicHulls.wl} package\cite{conichull}. The evaluation gives the following result:
\begin{align}\label{b3sol}
    B_{3}(s)&= -\frac{\pi }{2 \left(s^2+5 s+6\right)}+\frac{\sqrt{\pi } \left((s+2) \, _2F_1\left(\frac{1}{2},-\frac{s}{2}-\frac{1}{2};\frac{3}{2};-1\right)+\, _2F_1\left(-\frac{s}{2}-1,-\frac{s}{2}-\frac{1}{2};-\frac{s}{2};-1\right)\right) \Gamma \left(-\frac{s}{2}-\frac{1}{2}\right) \Gamma (s+2)}{2 (s+3) \Gamma \left(-\frac{s}{2}\right) \Gamma (s+3)}+ \nonumber \\
    & \frac{1}{1+s} F{}^{2:1:1}_{1:1:1}
  \left[
   \setlength{\arraycolsep}{0pt}
   \begin{array}{c@{{}:{}}c@{;{}}c}
  \dfrac{-1-s}{2},\dfrac{-s}{2} & \dfrac{1}{2} & \dfrac{1}{2} \\[1ex]
   \dfrac{1-s}{2},\dfrac{1}{2} & \dfrac{1}{2} & \linefill
   \end{array}
   \;\middle|\;
   -1,-1
 \right]
\end{align}
Where $F^{2:1:1}_{1:1:1}(x,y)$ is the KdF function which converges for $|\sqrt{x}|+|\sqrt{y}|<1$. So, to evaluate it at $(-1,-1)$, one needs its analytic continuations. In the \mt file \texttt{Box.nb}, we provide a systematic derivation of the analytic continuation for the same so that it converges at $(-1,1)$.\\
For general $B_{n}(s)$ we get the following MB-representation
\begin{equation}\label{bngeneral}
    B_{n}(s)= \frac{1}{\Gamma \left(-\frac{s}{2}\right)} \int\limits_{c_1 - i \infty} ^{c_1 + i \infty}\cdots \int\limits_{c_{n-1} - i \infty} ^{c_{n-1} + i \infty}\, \left(\prod_{p=1}^{n-1}\frac{\mathrm{d}z_{p}}{2 \pi i} \right) \frac{\left(\prod _{i=1}^{n-1} \Gamma \left(2 z_i+1\right)\right) \Gamma \left(s-2 \sum _{j=1}^{n-1} z_j+1\right) \Gamma \left(\sum _{k=1}^{n-1} z_k-\frac{s}{2}\right)}{ \left(\prod _{l=1}^{n-1} \Gamma \left(2 z_l+2\right)\right) \Gamma \left(s-2 \sum _{m=1}^{n-1} z_m+2\right)}
\end{equation}
Using the Eq. \eqref{bngeneral} we obtain following representation for $B_{4}(s)$
\begin{align}
     B_{4}(s,\alpha,\beta,\gamma)= \frac{1}{\Gamma \left(-\frac{s}{2}\right)}\int\limits_{c_1 - i \infty} ^{c_1 + i \infty}\, \int\limits_{c_2 - i \infty} ^{c_2 + i \infty}\, \int\limits_{c_3 - i \infty} ^{c_3 + i \infty}\, & \frac{\Gamma \left(-z_1\right) \Gamma \left(2 z_1+1\right) \Gamma \left(-z_2\right) \Gamma \left(2 z_2+1\right) \Gamma \left(-z_3\right) \Gamma \left(2 z_3+1\right) \Gamma \left(s-2 z_1-2 z_2-2 z_3+1\right) }{ \Gamma \left(2 z_1+2\right) \Gamma \left(2 z_2+2\right) \Gamma \left(2 z_3+2\right) \Gamma \left(s-2 z_1-2 z_2-2 z_3+2\right)} \nonumber \\
     & \times \Gamma \left(-\frac{s}{2}+z_1+z_2+z_3\right)(\alpha)^{z_1}(\beta)^{z_2}(\gamma)^{z_3}\frac{\mathrm{d}z_{1} \mathrm{d}z_{2} \mathrm{d}z_{3}}{(2 \pi i)^3} 
\end{align}
The above integral can be again evaluated readily using the \ch package. For the case of $B_{4}(s)$, due to the occurrence of a 3-variable hypergeometric function, the region of convergence analysis is difficult. In the OMOB, all the series which converge in the same region of convergence are kept together. For 3 or more variables, this analysis becomes complicated and is not always straightforward \cite{srivastava1985multiple}. Here, the CHMB method plays an important role in that it clubs the series converging in the same region of convergence together without prior knowledge of their region of convergence. The evaluation has been provided in the file \is.
\subsection{$\Delta_n(s)$}\label{subsection: Delta}
We now move on to the evaluation of $\Delta_n$ integrals \eqref{deltaint}. Instead of directly doing the evaluation of the $\Delta_{n}(s)$ integral, we refer to \cite{bailey2010advances}, to exploit the relation between $B_{n}(s)$ and $\Delta_{n}(s)$. A few instances of the same are as follows:
\begin{align}
    \Delta_1 (s) &= 2\frac{1}{(s+1)(s+2)}\\
    \Delta_2 (s) &= 8\frac{2^{\frac{s}{2}+1} (s+3)+1}{(s+2) (s+3) (s+4)}+4 B_2(s) -\frac{4 (s+4) }{s+2}B_2(s+2) \\
    \Delta_3 (s) &= 24 \frac{\left((s+5) \left(2^{\frac{s}{2}+3}-3^{\frac{s}{2}+2}\right)+1\right)}{(s+2) (s+4) (s+5) (s+6)} + \frac{24 }{s+2} B_2 (s+2)-\frac{24 (s+6)}{(s+2) (s+4)}B_2(s+4)  -\frac{12 (s+5)}{s+2}B_3(s+2)\\ \nonumber
    & \; +\frac{4 (s+6) (s+7)}{(s+2) (s+4)} B_3(s+4)+8 B_3(s)\\
\end{align}
where $B_{2}(s)$ and $B_{3}(s)$ are given by Eq.\eqref{b2sol} and Eq.\eqref{b3sol}.
The results for $\Delta_4$ and $\Delta_5$ are provided in the appendix \ref{section: Delta values}.
 
\subsection{Jellium Potential}
As an application of the evaluations done in the previous section, we refer to one more application of such evaluations, the Jellium potential\cite{BAILEY2007196}. It arises in the problem of electrostatics. The problem concerns finding the electrostatic potential energy of an electron (having charge -1) at the cube centre, given an $n$-cube of uniformly charged jelly of total charge $+1$. For the problem, usually  one takes the radial potential at a distance $r$ from the electron as $V_{n}(r)$ as follows
\begin{align} 
    V_{1}(r)&:= r-1/2, \nonumber \\
    V_{2}(r)&:= \log(2r), \nonumber \\
    V_n(r)&:=2^{n-2}-\left(\frac{1}{r}\right)^{n-2}, \quad n>2
\end{align}
The $n$-th Jellium potential is defined as
\begin{equation}
    J_n:=\left\langle V_n(r)\right\rangle_{\vec{r} \in[-1 / 2,1 / 2]^n}
\end{equation}
All the $J_n$ can be written as a box integral up to an offset. The final result is 
\begin{equation}
  J_n=2^{n-2}\left(1-B_n(2-n)\right), \quad n>2  
\end{equation}
Using the result for $B_n$, $J_3$ can be readily evaluated to:
\begin{equation}
    J_3 = \frac{\pi }{2}+2-6 \tanh ^{-1}\left(\frac{1}{\sqrt{3}}\right)
\end{equation}

\section{Conclusion and Discussion}\label{conclusions}
We show that using the MMOB\cite{prausa2017mellin} for the evaluation of improper integral with limits from $0$ to $\infty$ combined with tools to evaluate such MB integrals such as \ch results in more efficient evaluation of the integrals considered here. This method is particularly helpful to evaluate the integrals when using OMOB; one requires the use of `regulators' and a proper limiting procedure to evaluate these integrals. The choice of these regulators is somewhat arbitrary, and at times, more than one regulator has to be used, which further complicates the process. With these tools at hand, we then re-evaluate the Ising integral, which had been already evaluated in \cite{Gonzalez:2010nm} but with regulators. We further make an attempt to evaluate the sought-after integral $C_{5,k}$ with all these techniques. We are, though, able to evaluate a more general integral $C_{5,k}(\alpha,\beta)$, which, when properly analytically continued, will give the result for $C_{5,k}$. At present, we are unable to do so with the techniques at hand. However, we believe that the result can be written as a derivative of some multivariable hypergeometric function. Continuing further we evaluate the $B_{3}(s)$ and $B_{4}(s)$ and give a general MB representation for $B_{n}(s)$. For the case of $B_{3}(s)$, we use \ol to find the ACs of the hypergeometric functions that appear in the solution. For $B_{4}(s)$, similar techniques would work. It is important to note that though the OMOB and the evaluation of MB representation will give essentially the same number of series, grouping them in the same ROC is not an easy task. For the case of 3 or more variables, the problem of finding the ROC is still a problem yet to be solved in an efficient manner. This problem is essentially removed in the case of applying the CHMB method, where such grouping is automatically done without prior knowledge of the ROC. As a byproduct of these evaluations, we get the result for associated box integrals $\Delta_{n}(s)$ and Jellium potential $J_{n}$. Through these evaluations, we also discover the relations between these integrals and multivariable hypergeometric functions. We would like to emphasize again that the present work is an illustration of combining various techniques and automatizations \cite{Ananthanarayan:2021yar,conichull} together, which are developed as a part of a long series of investigations, rather than using them in isolation.

As a future direction, it would be interesting to modify the rules of the OMOB so that the final evaluation of the bracket series doesn't require regulators. For the case of $C_{5,k}(\alpha,\beta)$ evaluated in the present work, one can try to find a way to evaluate the ACs. One way towards this direction is to write the final result as a derivative of a hypergeometric function and then find the ACs of it using \ol. After finding the ACs, the derivative can be taken to get the final result, which converges in the appropriate region. We also note that a similar process can be used to evaluate $C_{6,k}$, which also gives a 2-fold MB integral. Finally, it would be interesting to derive the result for the various Box integrals $B_{n}(s), \Delta_{n}(s)$ and Jellium-potential $J_{n}$ from the results given here. The result in the present work matches numerically with those results; it would still be interesting to see how they can be obtained from the present work by using various reduction formulas of multivariable hypergeometric functions.  
\section{Acknowledgements}
TP would like to thank Souvik Bera for his help and useful comments.
\appendix
\section{Ruby's formula }\label{rubymob}
Ruby's formula is another interesting physical problem where the OMOB can still be used. We provide an evaluation of a general integral of which Ruby's formula is a special case in this Appendix to highlight the application of the OMOB when regulators are not required. Ruby's formula gives the solid angle subtended at a disk source by a coaxial parallel-disk detector \cite{RUBY1994531}. It is given as follows\\
\begin{equation}\label{ruby_formula}
D = \frac{R_d}{R_s}\int_{0}^{\infty} J_1(k R_d) J_1(k R_s) \frac{e^{-kd}}{k}\, \mathrm{d}k
\end{equation}
 \\where $R_d$ and $R_s$  are the radii of the detector and the source, respectively, $d$ is the distance between the source and the detector, and $J_{1}(x)$ is the order one Bessel's function of the first kind.
We now consider the generalization of integral \eqref{ruby_formula}, as discussed in \cite{Friot:2014ufa}. We will use the MOB to evaluate the integral and show that it reproduces the result, along with two ACs. 
\begin{equation}\label{28}
S = \int_{0}^{\infty} k^l e^{-kd}\prod_{j=1}^{N}J_{a_j}(kR_j) \, \mathrm{d} {k}
\end{equation}
we can again apply the method of brackets by using the series expansion of the functions \\
\begin{equation*}
J_{a_j}(kR_j)= \frac{1}{2^{a_j}}\sum_{n_j=0}^{\infty}\phi_{n_j} \frac{(kR_j)^{2n_j+a_j}}{2^{2n_j}\Gamma(a_j+n_j+1)}
\end{equation*}\\
\begin{equation*}
e^{-kd} = \sum_{n_p=0}^{\infty}\phi_{n_p}k^{n_p}  d^{n_p}
\end{equation*}
putting the series expansion in the above integral, we get\\
\begin{equation}\label{29}
S = \int_{0}^{\infty}\sum_{n_p=0}^{\infty}\phi_{n_p}k^{n_p+l}  d^{n_p}\prod_{j=1}^{N}\frac{1}{2^{a_j}}\sum_{n_j=0}^{\infty}\phi_{n_j} \frac{(kR_j)^{2n_j+1}}{2^{2n_j}\Gamma(a_j+n_j+1)} \mathrm{d}{k}
\end{equation}\\\\
we can simplify the above by noting that\\\\
\begin{equation*}
\begin{split}
\prod_{j=1}^{N}\frac{1}{2^{a_j}}\sum_{n_j=1}^{\infty}\phi_{n_j} \frac{(kR_j)^{2n_j+a_j}}{2^{2n_j}\Gamma(a_j+n_j+1)} = \sum_{n_1=0}^{\infty} \cdots \sum_{n_N=0}^{\infty}\frac{\phi_{1,2, \cdots, N}k^{\sum_{j=1}^{N}(2n_j+a_j)}}{2^{\sum_{j=1}^{N}(2n_j+a_j)}}\\
\times\frac{\prod_{j=1}^{N}(R_j)^{(2n_j+a_j)}}{\prod_{j=1}^{N}\Gamma(a_j+n_j+1)}
\end{split}
\end{equation*}\\\\
putting above value in Eq.\eqref{29}  gives\\
\begin{equation}\label{30}
\begin{split}
S = \int_{0}^{\infty}\sum_{n_p=0}^{\infty}\phi_{n_p}k^{(n_p+l+\sum_{j=1}^{N}(2n_j+a_j))}  d^{n_p}\sum_{n_1=0}^{\infty}\cdots\sum_{n_N=0}^{\infty}\frac{\phi_{1,2, \cdots ,N}}{2^{\sum_{j=1}^{N}(2n_j+a_j)}}\\
\times\frac{\prod_{j=1}^{N}(R_j)^{(2n_j+a_j)}}{\prod_{j=1}^{N}\Gamma(a_j+n_j+1)}\mathrm{d}{k}
\end{split}
\end{equation}\\
Using the method of brackets, Eq.\eqref{30} can be written as\\\\
\begin{equation}\label{31}
\begin{split}
S = \sum_{n_1=0}^{\infty}\cdots\sum_{n_N=0}^{\infty}\sum_{n_p=0}^{\infty}\phi_{1,2, \cdots ,N,p} \langle(n_p+l+1+\sum_{j=1}^{N}(2n_j+a_j))\rangle\frac{ d^{n_p}}{2^{\sum_{j=1}^{N}(2n_j+a_j)}}\\
\times\frac{\prod_{j=1}^{N}(R_j)^{(2n_j+a_j)}}{\prod_{j=1}^{N}\Gamma(a_j+n_j+1)}
\end{split}
\end{equation}
where \(\phi_{1, 2, \cdots, N, p}\) = \(\phi_{n_1}\phi_{n_2}\cdots\phi_{n_N}\phi_{n_p}\)\\
The solutions to Eq.\eqref{31}  are determined using the solution to the linear equation.
\begin{equation}\label{32}
n_p+l+1+\sum_{j=1}^{N}(2n_j+a_j)=0
\end{equation}\\
above equation has ($N+1$) variables.  There are ($N+1$) different ways to write solutions to the above equation, taking $N$ free variables each time.\\\\
Out of ($N$+1) solutions, the solution with \(n_p\) as the dependent variable gives the Lauricella function of $N$ variables, as we will show.  The rest of the solutions give the series representation that is the analytical continuation of the earlier.\\\\
Denoting the solution to Eq.\eqref{32}  by \(n_i^*\) with \(n_i\) being the dependent variable. The solutions to equation Eq.\eqref{32} can be written as
\begin{equation*}
n_p^*=-(l+1)-\sum_{j=1}^{N}(2n_j+a_j) ; a= 1
\end{equation*}\\
\begin{equation*}
n_i^*=-\frac{(n_p+l+1)}{2}-\sum_{j=1,i\neq j}^{N}(n_j)-\sum_{j=1}^{N}\Big(\frac{a_j}{2}\Big); a= \frac{1}{2}
\end{equation*}\\
$\boldsymbol{a}$ is the coefficient of the dependent variable if the set of linear equations obtained from brackets are written in the form \(\boldsymbol{an+b= 0}\), where $n$ is the dependent variable, and $b$ includes all the free variables and the constants.\\\\
Denoting the solution of Eq.\eqref{31}  by \(S_i\) obtained by using \(n_i^*\) $(i=1,2,\cdots,N,p)$.\\\\
i) \textbf {With \(n_p\) as the dependent variable}\\

We write the solution to Eq.\eqref{31} as\\\\
\begin{equation}
S_p = \frac{1}{a}\sum_{n_1=0}^{\infty}\cdots\sum_{n_N=0}^{\infty}\phi_{1,2,\cdots,N}F(n_1,n_2,\cdots ,n_N,n_p^*)\Gamma(-n_p^*)
\end{equation}\\\\
where \(F(n_1,n_2,\cdots,n_N,n_p)= \frac{ d^{n_p}\prod_{j=1}^{N}(R_j)^{(2n_j+a_j)}}{2^{\sum_{j=1}^{N}(2n_j+a_j)}\prod_{j=1}^{N}\Gamma(a_j+n_j+1)}\).\\\\\\
Putting the values, we get
\begin{equation} \label{35}
S_p = \sum_{n_1=0}^{\infty}\cdots\sum_{n_N=0}^{\infty} \phi_{1,2,\cdots,N} \frac{ d^{-(l+1)-\sum_{j=1}^{N}(2n_j+a_j)}\prod_{j=1}^{N}(R_j)^{(2n_j+a_j)}}{2^{\sum_{j=1}^{N}(2n_j+a_j)}\prod_{j=1}^{N}\Gamma(a_j+n_j+1)}
\Gamma\Big((l+1)+\sum_{j=1}^{N}(2n_j+a_j)\Big)
\end{equation}\\\\\\
Using Legendre's duplication formula\\\\
\begin{equation}
\Gamma\bigg(2\Big(\frac{l+1}{2}+\sum_{j=1}^{N}\Big(n_j+\frac{a_j}{2}\Big)\Big)\bigg) = \frac{2^{\Big(l+\sum_{j=1}^{N}(2n_j+a_j)\Big)}\Gamma\Big(\frac{l+1}{2}+\sum_{j=1}^{N}\Big(n_j+\frac{a_j}{2}\Big)\Big)\Gamma\Big(\frac{l}{2}+1+\sum_{j=1}^{N}\Big(n_j+\frac{a_j}{2}\Big)\Big)}{\sqrt{\pi}}
\end{equation}\\\\
putting above value in equation  Eq.\eqref{35}  and simplifying gives\\\\
\begin{equation}
\begin{split}
S_p = \sum_{n_1=0}^{\infty}\cdots\sum_{n_N=0}^{\infty}\phi_{1,2,\cdots,N} \frac{ d^{-(l+1)-\sum_{j=1}^{N}(2n_j+a_j)}\prod_{j=1}^{N}(R_j)^{(2n_j+a_j)}}{2^{\sum_{j=1}^{N}(2n_j+a_j)}\prod_{j=1}^{N}\Gamma(a_j+n_j+1)}\\
\times\frac{\Gamma\Big(\frac{l+1}{2}+\sum_{j=1}^{N}\Big(n_j+\frac{a_j}{2}\Big)\Big)\Gamma\Big(\frac{l}{2}+1+\sum_{j=1}^{N}\Big(n_j+\frac{a_j}{2}\Big)\Big)}{\sqrt{\pi}}
\end{split}
\end{equation}\\\\\\
this equation can be written in compact form  as follows \\\\
\begin{equation}
\begin{split}
S_p= \frac{1}{\sqrt{\pi}}\Big(\frac{2}{d}\Big)^{l}\Big(\frac{1}{d}\Big)\Gamma\Big(\sum_{j=1}^{N}\frac{a_j}{2}+\frac{l+1}{2}\Big)\Gamma\Big(\sum_{j=1}^{N}\frac{a_j}{2}+\frac{l}{2}+1\Big)\prod_{j=1}^{N}\Big(\frac{R_j}{d}\Big)^{a_j} \\
\times\sum_{n_1=0}^{\infty}\cdots\sum_{n_N=0}^{\infty}\frac{(-1)^{\sum_{j=1}^{N}n_j}\prod_{j=1}^{N}\Big(\frac{R_j}{d}\Big)^{2n_j}}{\prod_{j=1}^{N}\bigg(\Big(a_j+1\Big)_{n_j} \Gamma(n_j+1)\bigg)}\\
\times\frac{\bigg(\sum_{j=1}^{N}\frac{a_j}{2}+\frac{l+1}{2}\bigg)_{(\sum_{j=1}^{N}n_j)} \bigg(\sum_{j=1}^{N}\frac{a_j}{2}+\frac{l}{2}+1\bigg)_{(\sum_{j=1}^{N}n_j)}}{\prod_{j=1}^{N}\Gamma(a_j+1)}
\end{split}
\end{equation}\\
\((a)_m\) is the Pochhammer symbol\\\\
which exactly matches the series representation obtained in \cite{Friot:2014ufa} with ROC  \\\\
\centerline{$\displaystyle\sum_{i=1}^{N}| R_j| < d$}\\\\
The above series corresponds to the Lauricella function of $N$ variables.
\begin{equation}\label{lauri_Fc}
\begin{split}
S_p= \frac{1}{\sqrt{\pi}}\Big(\frac{2}{d}\Big)^{l}\Big(\frac{1}{d}\Big)\Big(\frac{1}{\prod_{j=1}^{N}\Gamma(a_j+1)}\Big)\Gamma\Big(\sum_{j=1}^{N}\frac{a_j}{2}+\frac{l+1}{2}\Big)\Gamma\Big(\sum_{j=1}^{N}\frac{a_j}{2}+\frac{l}{2}+1\Big)\prod_{j=1}^{N}\Big(\frac{R_j}{d}\Big)^{a_j} \\
\times F_c\Bigg(\Big(\sum_{j=1}^{N}\frac{a_j}{2}+\frac{l+1}{2}\Big) , \Big(\sum_{j=1}^{N}\frac{a_j}{2}+\frac{l}{2}+1\Big) ;(1+a_1),\cdots,(1+a_N);-\Big(\frac{R_1}{d}\Big)^2,\cdots,-\Big(\frac{R_N}{d}\Big)^2\Bigg)
\end{split}
\end{equation}\\\\
where \(F_c\) in the above equation is the Lauricella function for $N$ variables.\\\\
ii) \textbf {With \(n_i\) as the dependent variable}\\

We write the solution to Eq.\eqref{31} as\\
\begin{equation}
S_i = \frac{1}{a}\sum_{n_1=0}^{\infty}\cdots\sum_{n_N=0}^{\infty}\phi_{1,2,\cdots,(i-1),(i+1),\cdots,N,p}F(n_1,n_2,\cdots,n_i^*,\cdots,n_N,n_p)\Gamma(-n_i^*)
\end{equation}\\\\
putting the values, we get\\
\begin{equation}\label{laurfc_ac}
\begin{split}
S_i = \frac{1}{2}\sum_{n_1=0}^{\infty}\cdots\sum_{n_{i-1}=0}^{\infty}\sum_{n_{i+1}=0}^{\infty}\cdots\sum_{n_N=0}^{\infty}\sum_{n_p=0}^{\infty}\phi_{1,2,\cdots,(i-1),(i+1),\cdots,N,p} \frac{ d^{n_p}\Big(\prod_{j=1,j\neq i}^{N}(R_j)^{(2n_j+a_j)}\Big)}{\Big(2^{\sum_{j=1,j\neq i}^{N}(2n_j+a_j)}\Big)\Big(\prod_{j=1,j\neq i}^{N}\Gamma(a_j+n_j+1)\Big)}\\
\times\Bigg(\frac{1}{\left(\Gamma(a_i-\frac{(n_p+l+1)}{2}-\sum_{n_j=1,i\neq j}^{N}(n_j)-\sum_{j=1}^{N}\left(\frac{a_j}{2}\right)+1\right)}\Bigg)\bigg((R_i)^{-(n_p+l+1)-\sum_{j=1,i\neq j}^{N}(2n_j)-\sum_{j=1}^{N}a_j+a_i}\bigg)\\
\times\Bigg(\frac{1}{2^{-(n_p+l+1)-\sum_{j=1,i\neq j}^{N}(2n_j)-\sum_{j=1}^{N}a_j+a_i}}\Bigg)\bigg(\Gamma\Big(\frac{n_p+l+1}{2}+\sum_{j=1,j\neq i}^{N}(n_j)+\sum_{j=1}^{N}\left(\frac{a_j}{2}\right)\Big)\bigg)
\end{split}
\end{equation}\\
Eq.\eqref{laurfc_ac} gives series representation for all values of $i = 1, 2, \cdots, N$ and is the most general form of all the analytically continued series.

\section{$\Delta_n$ Relations}\label{section: Delta values}
$\Delta_n$ can be expressed in terms of $B_n$ as has already been shown in the subsection \eqref{subsection: Delta}. Here, the relations for $\Delta_4$ and $\Delta_5$ are provided:
\begin{align}
\Delta_4(s)&= 64 \frac{\left(3 \cdot 2^{\frac{s}{2}+3}+2^{s+6}-3^{\frac{s}{2}+4}\right)(s+7)+1}{(s+2)(s+4)(s+6)(s+7)(s+8)}+\frac{96}{(s+2)(s+4)} B_2(s+4)-\frac{96(s+8)}{(s+2)(s+4)(s+6)} B_2(s+6) \\
& +\frac{64}{s+2} B_3(s+2) -\frac{96(s+7)}{(s+2)(s+4)} B_3(s+4)+\frac{32(s+8)(s+9)}{(s+2)(s+4)(s+6)} B_3(s+6)+16 B_4(s) \nonumber\\
& -\frac{88(s+6)}{3(s+2)} B_4(s+2)+\frac{8(s+8)(6 s+43)}{3(s+2)(s+4)} B_4(s+4) -\frac{8(s+8)(s+9)(s+10)}{3(s+2)(s+4)(s+6)} B_4(s+6) \nonumber\\
\Delta_5(s)&=160 \frac{1+(9+s)\left(2^{6+s / 2}+2^{10+s}-5^{4+s / 2}-2 \cdot 3^{5+s / 2}\right)}{(2+s)(4+s)(6+s)(8+s)(9+s)(10+s)} +\frac{320}{(2+s)(4+s)(6+s)} B_2(6+s)+\frac{320}{(2+s)(4+s)} B_3(4+s)  \\
&  -\frac{320(10+s)}{(2+s)(4+s)(6+s)(8+s)} B_2(8+s)  -\frac{480(9+s)}{(2+s)(4+s)(6+s)} B_3(6+s) +\frac{160}{2+s} B_4(2+s) \nonumber\\
&  -\frac{880}{3} \frac{(8+s)}{(2+s)(4+s)} B_4(4+s)+\frac{80}{3} \frac{(10+s)(55+6 s)}{(2+s)(4+s)(6+s)} B_4(6+s)-\frac{80}{3} \frac{(10+s)(11+s)(12+s)}{(2+s)(4+s)(6+s)(8+s)} B_4(8+s)  \nonumber\\
& +32 B_5(s)-200 \frac{(7+s)}{6+3 s} B_5(2+s)+\frac{4}{3} \frac{(9+s)(291+35 s)}{(2+s)(4+s)} B_5(4+s)-\frac{8}{3} \frac{(10+s)(11+s)(47+5 s)}{(2+s)(4+s)(6+s)} B_5(6+s)  \nonumber\\
& +\frac{4}{3} \frac{(10+s)(11+s)(12+s)(13+s)}{(2+s)(4+s)(6+s)(8+s)} B_5(8+s)  \nonumber
\end{align}
\section{\mt files}\label{mfiles}
Here, we give a list of the \mt files and packages that we provide, which contains the derivation of the various results of the paper.
\begin{table}[h]
\centering
\begin{tabular}{|l|l|}
\hline
 \textbf{Files Provided} & \textbf{Description} \\ \hline
 \is & Contains the evaluation of the Ising integrals $C_{3,k}$, $C_{4,k}$, $C_{5,k}(\alpha,\beta)$ and $C_{6,k}(\alpha,\beta)$ \\ \hline
 \bo & Contains the evaluation of the Box integrals $B_{3}(s)$ and $B_{4}(s)$  \\ \hline
 \ch & Package required to evaluate multidimensional MB integrals. Used in the   \\
 & evaluation of $C_{5,k}(\alpha,\beta)$, $C_{6,k}(\alpha,\beta)$, $B_{3}(s)$ and $B_{4}(s)$ \\ \hline
 \texttt{MultivariateResidues.m} & Used by the package \ch internally  \\ \hline
 \ol & Package required for finding the ACs. Used for the case $B_{3}(s)$  \\ \hline
 \texttt{ROC2.wl} & Package required for finding the region of convergence of the 2-variable hypergeometric series. \\ \hline
\end{tabular}
\caption{}
\label{mtfiles}
\end{table}
\section{Conflict of Interest Statement}
We have no conflicts of interest to disclose.
\section{Data Availability Statement}
Data sharing is not applicable to this article as no datasets were generated or analyzed during the current study.
\bibliographystyle{unsrt}
\bibliography{reference}

\begin{thebibliography}{10}

\bibitem{Smirnov:2006ry}
V.~A. Smirnov.
\newblock {\em {Feynman integral calculus}}.
\newblock Springer, Berlin, Heidelberg, 2006.

\bibitem{Dubovyk:2022obc}
Ievgen Dubovyk, Janusz Gluza, and Gabor Somogyi.
\newblock {Mellin-Barnes Integrals: A Primer on Particle Physics Applications}.
\newblock {\em Lect. Notes Phys.}, 1008:pp., 2022.

\bibitem{Ananthanarayan:2020ncn}
B.~Ananthanarayan, Sumit Banik, Samuel Friot, and Shayan Ghosh.
\newblock {Double box and hexagon conformal Feynman integrals}.
\newblock {\em Phys. Rev. D}, 102(9):091901, 2020.

\bibitem{Ananthanarayan:2020xpd}
B.~Ananthanarayan, Sumit Banik, Samuel Friot, and Shayan Ghosh.
\newblock {Massive One-loop Conformal Feynman Integrals and Quadratic
  Transformations of Multiple Hypergeometric Series}.
\newblock {\em Phys. Rev. D}, 103(9):096008, 2021.

\bibitem{Banik:2022quy}
Sumit Banik.
\newblock {\em {On Hypergeometric solutions of Feynman integrals using
  Mellin-Barnes Integrals with Applications}}.
\newblock PhD thesis, Bangalore, Indian Inst. Sci., 9 2022.

\bibitem{hardy1999ramanujan}
Godfrey~Harold Hardy.
\newblock {\em Ramanujan: Twelve lectures on subjects suggested by his life and
  work}, volume 136.
\newblock American Mathematical Soc., 1999.

\bibitem{GONZALEZ201050}
Ivan Gonzalez and Victor~H. Moll.
\newblock Definite integrals by the method of brackets.
\newblock {\em Advances in Applied Mathematics}, 45(1):50--73, 2010.

\bibitem{Gonzalez:2010nm}
Ivan Gonzalez, Victor~H. Moll, and Armin Straub.
\newblock {The Method of brackets. Part 2. Examples and applications}.
\newblock 4 2010.

\bibitem{GonzalezKohlJiuMoll1}
Ivan Gonzalez, Karen Kohl, Lin Jiu, and Victor~H. Moll.
\newblock An extension of the method of brackets. part 1.
\newblock {\em Open Mathematics}, 15(1):1181--1211, 2017.

\bibitem{quadraticmob}
B.~Ananthanarayan, Sumit Banik, Sudeepan Datta, and Tanay Pathak.
\newblock {Quadratic and quartic integrals using the method of brackets}.
\newblock {\em Scientia}, 29:45--59, 2019.

\bibitem{GonzalezJiuMoll2}
Ivan Gonzalez, Lin Jiu, and Victor~H. Moll.
\newblock An extension of the method of brackets. part 2.
\newblock {\em Open Mathematics}, 18(1):983--995, 2020.

\bibitem{gonzalez2022mellin}
Ivan Gonzalez, Igor Kondrashuk, Victor~H Moll, and Luis~M Recabarren.
\newblock Mellin--barnes integrals and the method of brackets.
\newblock {\em The European Physical Journal C}, 82(1):28, 2022.

\bibitem{gonzalez2022analytic}
Ivan Gonzalez, Igor Kondrashuk, Victor~H Moll, and Alfredo Vega.
\newblock Analytic expressions for debye functions and the heat capacity of a
  solid.
\newblock {\em Mathematics}, 10(10):1745, 2022.

\bibitem{Ananthanarayan:2021not}
B.~Ananthanarayan, Sumit Banik, Samuel Friot, and Tanay Pathak.
\newblock {Method of Brackets: Revisiting a technique for calculating Feynman
  integrals and certain definite integrals}.
\newblock {\em Phys. Rev. D}, 108(8):085001, 2023.

\bibitem{gradshteyn2014table}
Izrail~Solomonovich Gradshteyn and Iosif~Moiseevich Ryzhik.
\newblock {\em Table of integrals, series, and products}.
\newblock Academic press, 2014.

\bibitem{appellf2}
B.~Ananthanarayan, Souvik Bera, S.~Friot, O.~Marichev, and Tanay Pathak.
\newblock {On the evaluation of the Appell F2 double hypergeometric function}.
\newblock {\em Comput. Phys. Commun.}, 284:108589, 2023.

\bibitem{Ananthanarayan:2021yar}
B.~Ananthanarayan, Souvik Bera, S.~Friot, and Tanay Pathak.
\newblock {Olsson.wl : a $Mathematica$ package for the computation of linear
  transformations of multivariable hypergeometric functions}.
\newblock 12 2021.

\bibitem{Bera:2022eag}
Souvik Bera and Tanay Pathak.
\newblock {Analytic continuations of the Horn $H_1$ and $H_5$ functions}.
\newblock 10 2022.

\bibitem{srivastava1985multiple}
Hari~M Srivastava and Per~Wennerberg Karlsson.
\newblock {\em Multiple Gaussian hypergeometric series}.
\newblock E. Horwood, 1985.

\bibitem{exton1976multiple}
Harold Exton.
\newblock {\em Multiple hypergeometric functions and applications}.
\newblock Ellis Horwood, 1976.

\bibitem{conichull}
B.~Ananthanarayan, Sumit Banik, Samuel Friot, and Shayan Ghosh.
\newblock {Multiple Series Representations of N-fold Mellin-Barnes Integrals}.
\newblock {\em Phys. Rev. Lett.}, 127(15):151601, 2021.

\bibitem{Banik:2022bmk}
Sumit Banik and Samuel Friot.
\newblock {Multiple Mellin-Barnes integrals with straight contours}.
\newblock {\em Phys. Rev. D}, 107(1):016007, 2023.

\bibitem{Olsson64}
{Olsson, Per O. M. }.
\newblock {Integration of the Partial Differential Equations for the
  Hypergeometric Functions $F_1$ and $F_D$ of Two and More Variables}.
\newblock {\em {Journal of Mathematical Physics}}, 5(3):420--430, 1964.

\bibitem{BAILEY2011741}
D.H. Bailey and J.M. Borwein.
\newblock High-precision numerical integration: Progress and challenges.
\newblock {\em Journal of Symbolic Computation}, 46(7):741--754, 2011.
\newblock Special Issue in Honour of Keith Geddes on his 60th Birthday.

\bibitem{bailey2006integrals}
David~H Bailey, Jonathan~M Borwein, and Richard~E Crandall.
\newblock Integrals of the ising class.
\newblock {\em Journal of Physics A: Mathematical and General}, 39(40):12271,
  2006.

\bibitem{stan2010recurrences}
Flavia Stan.
\newblock On recurrences for ising integrals.
\newblock {\em Advances in Applied Mathematics}, 45(3):334--345, 2010.

\bibitem{bailey2007hypergeometric}
David~H Bailey, David Borwein, Jonathan~M Borwein, and Richard~E Crandall.
\newblock Hypergeometric forms for ising-class integrals.
\newblock {\em Experimental Mathematics}, 16(3):257--276, 2007.

\bibitem{mobprob}
B.~Ananthanarayan, Sumit Banik, Samuel Friot, and Tanay Pathak.
\newblock {On the Method of Brackets}.
\newblock 12 2021.

\bibitem{prausa2017mellin}
Mario Prausa.
\newblock Mellin--barnes meets method of brackets: a novel approach to
  mellin--barnes representations of feynman integrals.
\newblock {\em The European Physical Journal C}, 77(9):1--10, 2017.

\bibitem{bailey2007box}
David~H Bailey, Jonathan~M Borwein, and Richard~E Crandall.
\newblock Box integrals.
\newblock {\em Journal of Computational and Applied Mathematics},
  206(1):196--208, 2007.

\bibitem{bailey2010advances}
D~Bailey, J~Borwein, and R~Crandall.
\newblock Advances in the theory of box integrals.
\newblock {\em Mathematics of Computation}, 79(271):1839--1866, 2010.

\bibitem{doi:10.1137/0130003}
R.~S. Anderssen, R.~P. Brent, D.~J. Daley, and P.~A.~P. Moran.
\newblock Concerning $\int_0^1 \cdots \int_0^1 {(x_1^2 + \cdots + x_k^2 )} ^{{1
  / 2}} dx_1 \cdots ,dx_k $ and a taylor series method.
\newblock {\em SIAM Journal on Applied Mathematics}, 30(1):22--30, 1976.

\bibitem{philip2008distance}
Johan Philip.
\newblock {\em The distance between two random points in a 4-and 5-cube}.
\newblock KTH mathematics, 2008.

\bibitem{BAILEY2007196}
D.H. Bailey, J.M. Borwein, and R.E. Crandall.
\newblock Box integrals.
\newblock {\em Journal of Computational and Applied Mathematics},
  206(1):196--208, 2007.

\bibitem{orrick2001susceptibility}
WP~Orrick, Bernie Nickel, AJ~Guttmann, and Jacques~HH Perk.
\newblock The susceptibility of the square lattice ising model: new
  developments.
\newblock {\em Journal of Statistical Physics}, 102:795--841, 2001.

\bibitem{PhysRevB.13.316}
Tai~Tsun Wu, Barry~M. McCoy, Craig~A. Tracy, and Eytan Barouch.
\newblock Spin-spin correlation functions for the two-dimensional ising model:
  Exact theory in the scaling region.
\newblock {\em Phys. Rev. B}, 13:316--374, Jan 1976.

\bibitem{zenine2005square}
N~Zenine, S~Boukraa, S~Hassani, and JM~Maillard.
\newblock Square lattice ising model susceptibility: series expansion method
  and differential equation for $\chi$ (3).
\newblock {\em Journal of Physics A: Mathematical and General}, 38(9):1875,
  2005.

\bibitem{bateman1953higher}
Harry Bateman.
\newblock {\em Higher transcendental functions [volumes i-iii]}, volume~1.
\newblock McGRAW-HILL book company, 1953.

\bibitem{bailey2010experimental}
David~H Bailey, Jonathan~M Borwein, David Broadhurst, and Wadim Zudilin.
\newblock Experimental mathematics and mathematical physics.
\newblock {\em Contemp. Math}, 517:41--58, 2010.

\bibitem{bailey2011high}
David~H Bailey and Jonathan~M Borwein.
\newblock High-precision numerical integration: Progress and challenges.
\newblock {\em Journal of Symbolic Computation}, 46(7):741--754, 2011.

\bibitem{olsson1964integration}
Per~OM Olsson.
\newblock Integration of the partial differential equations for the
  hypergeometric functions f 1 and fd of two and more variables.
\newblock {\em Journal of Mathematical Physics}, 5(3):420--430, 1964.

\bibitem{RUBY1994531}
Lawrence Ruby.
\newblock Further comments on the geometrical efficiency of a parallel-disk
  source and detector system.
\newblock {\em Nuclear Instruments and Methods in Physics Research Section A:
  Accelerators, Spectrometers, Detectors and Associated Equipment},
  337(2):531--533, 1994.

\bibitem{Friot:2014ufa}
Samuel Friot.
\newblock {On Ruby{'}s solid angle formula and some of its generalizations}.
\newblock {\em Nucl. Instrum. Meth. A}, 773:150--153, 2015.

\end{thebibliography}

\end{document}